\documentclass[prl,nobalancelastpage,twocolumn,nofootinbib,superscriptaddress,showpacs]{revtex4-1}
\usepackage{color,amsthm,amsmath,amsfonts,graphicx,bm}

\newcommand{\be}{\begin{equation}}
\newcommand{\ee}{\end{equation}}
\newcommand{\bea}{\begin{eqnarray}}
\newcommand{\eea}{\end{eqnarray}}

\newcommand{\tr}{\mathrm{tr}}

\newcommand{\dg}{\dagger}

\newcommand{\mb}{\mathbf}

\newcommand{\bs}{\boldsymbol}
\newcommand{\I}{\mathbb{I}}

\begin{document}

\title{Projected entangled-pair states can describe chiral topological states}

\author{T. B. Wahl}
\affiliation{Max-Planck Institut f\"ur Quantenoptik, Hans-Kopfermann-Str. 1, D-85748 Garching, Germany}

\author{H.-H. Tu}
\affiliation{Max-Planck Institut f\"ur Quantenoptik, Hans-Kopfermann-Str. 1, D-85748 Garching, Germany}

\author{N. Schuch}\affiliation{Institut f\"ur Quanteninformation, RWTH Aachen, D-52056 Aachen, Germany}

\author{J. I. Cirac}
\affiliation{Max-Planck Institut f\"ur Quantenoptik, Hans-Kopfermann-Str. 1, D-85748 Garching, Germany}

\date{\today}

\pacs{71.10.Hf, 73.43.-f}

\begin{abstract}
We show that Projected Entangled-Pair States (PEPS) in two spatial
dimensions can describe chiral topological states by explicitly
constructing a family of such states with a non-trivial Chern number. They
are ground states of two different kinds of free-fermion Hamiltonians: (i)
local and gapless; (ii) gapped, but with hopping amplitudes that decay
according to a power law. We derive general conditions on topological
free fermionic PEPS which show that they cannot correspond to exact
ground states of gapped, local parent Hamiltonians, and provide numerical
evidence demonstrating that they can nevertheless approximate well the
physical properties of topological insulators with local Hamiltonians at
arbitrary temperatures.
\end{abstract}

\maketitle

\emph{Introduction.---}%
Projected Entangled-Pair States (PEPS) \cite{Ver04} are believed to
provide an accurate description of many-body quantum systems with local
interactions in thermal equilibrium \cite{Has05}. At zero temperature,
PEPS contain the necessary entanglement demanded by the area law in order
to describe gapped Hamiltonians with local (short-range) interactions
\cite{Wol08}. In fact, Matrix-Product States, the 1D version of PEPS, have
been a key tool leading to the classification of all possible phases of
spin Hamiltonians of that kind \cite{Pol09,Che11,Sch11}.  Furthermore,
well known topological states in 2D, like the toric code \cite{Kit03},
resonating valence-bond \cite{And73}, or string-nets \cite{Lev05}, possess
a simple and exact description within that family
\cite{Ver06,Sch10,Bue08,Gu08}.  This seems to indicate that PEPS can also
help us to characterize and classify the gapped (topological) phases in
dimensions higher than one.

In spite of the above indications, there exists a deep reason to believe that
PEPS cannot describe the physics of certain kinds of topological phases,
namely those that have chirality. In fact, despite a significant effort in
the research of tensor network states, we do not know any PEPS
corresponding to a 2D chiral topological phase, not even for the simplest
topological insulators \cite{Note_Ber11, TopIns}. Those are free-fermionic systems
with a non-trivial Chern number, $C\ne 0$, that can be thought of as the
lattice counterpart of integer Quantum Hall materials. This fact may be
qualitatively understood as follows. Any PEPS is the ground state of a
local so-called \emph{parent} Hamiltonian $H=\sum_i h_i$ \cite{Per10,Cir09}.
This Hamiltonian is frustration-free, meaning that the PEPS is
annihilated by each local term $h_i$ individually.  But if $H$ corresponds
to a free-fermion system with a gapped band structure, the Hamiltonian
terms are of the form $h_i=b_i^\dagger b_i^{\phantom\dagger}$, where the
$b_i$ are quasiparticle operators supported in a small region. Their
Wannier functions must thus be localized, which in turn has been proven to
be impossible for systems with a non-trivial Chern number
\cite{Bro07,Mar12}.  Still, the question whether PEPS can describe chiral
topological insulators is open: First, even though their parent
Hamiltonians would be gapless, they might still be ground states of other
non-frustration-free gapped Hamiltonians (with non-localized Wannier
functions); second, although they do not provide exact descriptions of all
chiral states, they may still be able to approximate them accurately.

In this Letter, we explicitly construct a simple family of PEPS with Chern
number $C\ne 0$ on a square lattice. Our construction is based on Gaussian
fermionic PEPS (GFPEPS) \cite{Kra09}; that is, those that can be created
out of the vacuum by applying a Gaussian function of creation and
annihilation operators.  By simple we mean with the smallest possible bond
dimension, i.e., where there are just four auxiliary fermions on each
lattice site.  This family of GFPEPS possesses correlation functions with
a power-law decay, as well as non-localized Wannier functions. In fact,
they are the unique ground states of free-fermion, gapped Hamiltonians
with hopping amplitudes following the same decay. Apart from that, as all
PEPS, they are ground states of local parent Hamiltonians, which however
must be gapless due to the presence of critical correlations.  Indeed, we
prove that there cannot be GFPEPS with a non-trivial Chern number which
have a finite-range \emph{and} gapped parent Hamiltonian, since all such
Hamiltonians are in the trivial phase. This result, however, does not rule
out the possibility of using PEPS to {\em approximate} the ground state of
a chiral topological insulator with a local Hamiltonian. In fact, we
investigate this issue and conclude that this is possible, since the
approximation improves exponentially in the number of fermionic modes in
the bond.  Finally, we show that by using mixed GFPEPS we can approximate
the finite temperature properties of such systems as well.

\textit{Gaussian fermionic PEPS.---}We start with an $N\times N$ square
lattice with periodic boundaries and $f$ \textit{physical} fermionic orbitals at each site,
with creation (annihilation) operators $a_{\mathbf{r},j }^{\dagger }$
($a_{\mathbf{r},j }$), with $\mathbf{r}=(x,y)$ the site and $j =1, \ldots,
f$ the orbital index;  we will mostly work in the basis of physical
Majorana operators $c_{\mathbf{r},2j-1}=a_{\mathbf{r},j }^{\dagger
}+a_{\mathbf{r},j }$ and $c_{\mathbf{r},2j}=(-i)(a_{\mathbf{r},j
}^{\dagger }-a_{\mathbf{r},j })$.  To obtain a PEPS description of the
system, we start out with maximally entangled \emph{virtual} Majorana
modes $\gamma_{\mathbf{r},\alpha}^v$ (with $\alpha=1,\dots,\chi$ and
$v=l,r,u,d$) which are obtained by acting with $1 + i
\gamma_{\mathbf{r},\alpha}^r \gamma_{(\mathbf{r}+(1,0)),\alpha}^l$ and $1 + i
\gamma_{\mathbf{r},\alpha}^u \gamma_{(\mathbf{r}+(0,1)),\alpha}^d$ on the
vacuum (see Fig.~\ref{gfpeps}), yielding a pure state
$\rho_\mathrm{in}$~\cite{Kra09}. Here, the number of Majorana bonds $\chi$
is a parameter which can be used to systematically enlarge the class of
states. Subsequently, we apply the same linear map $\cal{E}$ to
each lattice site $\mathbf r$, which maps the $4\chi$ auxiliary modes
$\gamma_{\mathbf r,\alpha}^v$ to the $2f$ physical modes $c_{\mathbf
r,s}$ ($s = 1, \ldots, 2f$); this yields the translationally invariant fermionic PEPS
$\rho_{\mathrm{out}}$.

\begin{figure}
\centering\includegraphics[width=\columnwidth]{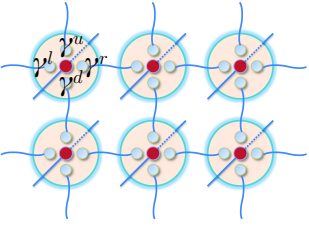}
\caption{Schematic of a GFPEPS in two dimensions. The Majorana modes (small gray balls)
form virtual bonds indicated by blue lines, which are mapped to the physical fermions
(red balls) by a Gaussian map denoted by big blue circles.}
\label{gfpeps}
\end{figure}

We now restrict to the case where the map ${\cal E}$ is Gaussian.  Then,
$\rho_\mathrm{out}$ is a free-fermion state, which can be described in
terms of its \emph{covariance matrix} (CM) $\Gamma_{\mathrm{out}}$,
defined as $(\Gamma_{\rm out})_{(\mathbf{r},s),(\mathbf{r}',t)} =
\frac{i}{2}{\rm tr}(\rho_{\rm out} [c_{\mathbf{r},s},c_{\mathbf{r}',t}])$;
similarly, for $\rho_\mathrm{in}$ we have $(\Gamma_{\rm
in})_{(\mathbf{r},\alpha),(\mathbf{r}',\beta)}^{v,v'} = \frac{i}{2}{\rm
tr}(\rho_{\rm in}
[\gamma_{\mathbf{r},\alpha}^v,\gamma_{\mathbf{r}',\beta}^{v'}])$.
Finally, $\mathcal E$ can be expressed using a CM $M$ defined on the
$2f+4\chi$ modes $\{(c_{\mathbf r,s}),(\gamma_{\mathbf r,\alpha}^v)\}$
which encodes how $\mathcal E$ correlates the input modes with the output
modes~\cite{Bravyi-2005} (an explicit expression will be given soon).
These CMs are real, antisymmetric, and fulfill $\Gamma \Gamma^{\top } \le
\I$, where equality only holds for pure states (and purity-preserving
maps).

Since we consider a translational invariant system of free fermions, it is
most convenient to work in Fourier space. The Fourier transformed CM
$G_{\mathrm{in}}$ of $\rho_\mathrm{in}$, expressed in terms of Fourier
transformed Majorana modes $\hat\gamma_{\mathbf{k},\alpha}^v =
\tfrac{1}{N} \sum_{\mb r} e^{- i \mb k \cdot \mb r}
\gamma_{\mathbf{r},\alpha}^v$ [$\mathbf{k}=(k_x,k_y)$,
$k_{x(y)}/(2\pi/N)=0,\ldots,N-1$], then reads
\begin{equation}
\label{eq:Gin}
G_{\mathrm{in}} (\mathbf{k}) = \left(\begin{array}{cc}
0& e^{i k_x}\,\I_{\chi}\\
- e^{-i k_x}\,\I_{\chi} &0\\
\end{array} \right) \oplus \left(\begin{array}{cc}
0& e^{i k_y}\,\I_{\chi}\\
- e^{-i k_y}\,\I_{\chi} &0\\
\end{array} \right)\ ,
\end{equation}
where $\I_\chi$ denotes a $\chi \times \chi$ identity matrix; the ordering
of the modes is $l,r,u,d$.  The CM $M$ for the Gaussian map $\mathcal E$
has a block structure
\begin{equation} \label{eq:M}
M=\left(
\begin{array}{cc}
A & B \\
-B^{\top } & D%
\end{array}%
\right) =-M^{\top },
\end{equation}%
where $A \in \mathbb{R}^{2 f \times 2 f}$, $B \in \mathbb{R}^{2 f \times 4
\chi}$, and $D \in \mathbb{R}^{4 \chi \times 4 \chi}$ are variational parameters corresponding to
physical and virtual modes. Any $M$ with $MM^\top\le \I$ characterizes an
admissible $\mathcal E$. Applying $\mathcal E$ to the input
$\rho_\mathrm{in}$ results in a CM~\cite{Bravyi-2005,Supp}
\begin{equation}
G_{\mathrm{out}} (\mathbf{k}) = {B} ({D} -
G_{\mathrm{in}}(\mathbf{k}))^{-1} {B}^{\top} + {A}\ \label{G_out}\ ,
\end{equation}
expressed in terms of the Fourier transformed physical Majorana modes $d_{\mathbf{k},s} = \tfrac{1}{N}
\sum_{\mb r} e^{- i \mb k \cdot \mb r} c_{\mathbf{r},s}$; $G_\mathrm{out}$
is pure if $MM^\top=\I$ (i.e., $\mathcal E$ preserves
purity).  Expressing the inverse in \eqref{G_out} by the adjugate matrix,
one finds that $\lbrack G
_{\mathrm{out}}(\mathbf{k})]_{st}=\frac{p_{st}(\mathbf{k})}{q(\mathbf{k})}$,
{where $p_{st}(\mathbf{k})$ and $q(\mathbf{k}) = \det(D - G_\mathrm{in}(\mb k))$ are trigonometric
polynomials of degree $\leq 2 \chi$ \cite{Supp}.}

For pure GFPEPS, the class of quadratic Hamiltonians
\begin{equation}
H_f=-i\sum_{\mathbf{k}} \sum_{s,t} \varepsilon(\mathbf{k}) \lbrack G _{\mathrm{out}}(\mathbf{k})]_{st} d_{\mathbf{k},s} d^\dg_{\mathbf{k},t},  \label{parent_Ham}
\end{equation}%
with spectrum $\varepsilon(\mb k) = {\varepsilon(- \mb k)}\ge0$ has
$\rho_\mathrm{out}$ as its ground state. {These \emph{parent Hamiltonians}
can have different properties: \emph{(i)} If $q(\mathbf k)>0$, then for
$\varepsilon(\mathbf{k})\equiv 1$, $H_f$ has exponentially decaying
two-body interactions in real space, and by choosing
$\varepsilon(\mathbf{k})=q(\mathbf k)$, one obtains a strictly local
gapped Hamiltonian.} \emph{(ii)} If $q(\mathbf{k})=0$ for some $\mathbf k$
and $G _{\mathrm{out}}(\mathbf{k})$ is continuous,
$\varepsilon(\mathbf{k})\equiv 1$ still yields a gapped Hamiltonian. Then,
whether $H_f$ has exponentially decaying terms depends on whether
$G_{\mathrm{out}}(\mathbf{k})$ has any discontinuities in its derivatives
(which give rise to algebraically decaying terms in real space after
Fourier transforming).

\textit{Example of a chiral GFPEPS.---}
Using this construction, we have obtained a family of chiral
topological insulators whose ground states are GFPEPS. They have $f=2$,
$\chi = 2$, and $M$ [Eq.~(\ref{eq:M})] is given by
\begin{align}
\begin{split}
A &= (-1 + 2 \lambda) \left(\begin{array}{cc}
\omega&0\\
0&-\omega
\end{array}\right) \\
B &= \sqrt{\frac{\lambda - \lambda^2}{2}} \left(\begin{matrix}
\I - \omega&\I + \omega&-\sqrt{2}\,\omega&\sqrt{2}\,\I\\
\I - \omega&-\I - \omega&\sqrt{2}\,\I&-\sqrt{2}\,\omega
\end{matrix}\right) \\
D &= \left(\begin{matrix}
0&(-1+\lambda)\,\I&-\frac{\lambda}{\sqrt{2}}\,\I&\frac{\lambda}{\sqrt{2}}\,\I\\\
(1-\lambda)\,\I&0&-\frac{\lambda}{\sqrt{2}}\,\I&-\frac{\lambda}{\sqrt{2}}\,\I\\
\frac{\lambda}{\sqrt{2}}\,\I&\frac{\lambda}{\sqrt{2}}\,\I&0&(-1+\lambda)\,\I\\
-\frac{\lambda}{\sqrt{2}}\,\I&\frac{\lambda}{\sqrt{2}}\,\I&(1-\lambda)\,\I&0
\end{matrix}\right)
\end{split}
\label{analytical_GFPEPS}
\end{align}
where $\mathbb I= \left(\begin{smallmatrix} 1&0\\  0&1
\end{smallmatrix}\right)$ and
$\omega = \left(\begin{smallmatrix} 0&1\\ -1&0
\end{smallmatrix}\right)$. The ordering of the physical Majorana modes is
$(c_{1\uparrow},c_{2\uparrow},c_{1\downarrow},c_{2\downarrow})$, and
that of the virtual modes as in (\ref{eq:Gin}); here, $0 <\lambda <
1$.
Using Eq.~(\ref{G_out}), one finds that
{
$G_{\mathrm{out}}(\mathbf k) = \frac{1}{\tilde q(\mathbf k)} \left(\begin{smallmatrix}
p_1(\mathbf k)\,\omega& i[p_3(\mathbf k)\,\I - p_2(\mathbf k)\,\omega]\\
i[p_3(\mathbf k)\,\I + p_2(\mathbf k)\,\omega]& - p_1(\mathbf k)\,\omega\\
\end{smallmatrix}\right)$
with $p_1(\mathbf k) = -2(1+\cos k_x)(1+\cos k_y )(1 - 2 \lambda) -
\lambda^2(1 + 2 \cos k_y + \cos k_x (2+3\cos k_y))$,
$p_2(\mathbf k) = 2(-\lambda + \lambda^2)(1 + \cos k_y )\sin k_x$, $p_3(\mathbf k) = 2(\lambda - \lambda^2)(1 + \cos k_x )\sin k_y $ and $\tilde q(\mathbf k)= 2(1+\cos k_x )(1+\cos k_y )(1 - 2 \lambda) + \lambda^2(3
+ 2 \cos k_y  + \cos k_x (2+\cos k_y))$}; note that
$G_\mathrm{out}(\mb k)$ is continuous but non-analytic at $\mathbf k = (\pi, \pi)$.

Employing Eq. \eqref{parent_Ham}, we can now define particle-number conserving
parent Hamiltonians for $G_\mathrm{out}$: If we choose $\varepsilon(\mb k) \equiv
1$, we obtain a gapped flat-band Hamiltonian with algebraically decaying
hoppings (see Fig.~\ref{analytical_gfpeps}), while if we choose
$\varepsilon(\mb k) = q(\mb k)$, we obtain a strictly local Hamiltonian with only
next-nearest neighbor couplings, 
which however is gapless
at $\mathbf k = (\pi, \pi)$ (inset of
Fig.~\ref{analytical_gfpeps}). In the first case, the Chern number can be
computed from $G_\mathrm{out}$ \cite{Supp} and is found to be $C =
-1$ for all $0 < \lambda < 1$.
Note that by changing to the basis $c_{1\uparrow}\pm c_{2\downarrow}$,
$c_{2\uparrow}\pm c_{1\downarrow}$, $G_\mathrm{out}$ decouples into two
GFPEPS describing spinless topological superconductors, each with $\chi=1$ and equal chiralities.

\textit{Conditions for topological GFPEPS.---}Let us now show that
topological GFPEPS are very special.  In particular, we will prove that any
GFPEPS with a property known as injectivity~\cite{Per08} (which holds
generically), or more generally for which $q(\mathbf{k})$ is
non-singular, has a gapped local parent Hamiltonian which is connected to
a trivial state via a gapped path and therefore cannot be topological;
{this implies that the parent Hamiltonians defined via $\epsilon(\mb k) = q(\mb k)$ have to be gapless.} (This shows that injectivity in GFPEPS is
much stronger than for general PEPS, where it does not have implications
about the spectrum or the phase except for 1D systems.)

Let us first define injectivity for GFPEPS: By blocking $n_v \times n_h$
sites to a new super-site (by tracing over the virtual particles), we can
reach a point where the number of physical Majorana modes $d_{\mathrm{ph}}
= 2f n_h n_v$ is larger than the number of virtual modes $d_{\mathrm{vir}}
= 2 \chi (n_h + n_v)$. Then, $G_{\mathrm{out} \Box}(\mathbf k) = B_{\Box}
(D_\Box - G_\mathrm{in \Box}(\mathbf k))^{-1} B_\Box^\top + A_\Box, $
where $\Box$ denotes the corresponding matrices after blocking. We say
that a GFPEPS is \emph{injective} if there is a finite blocking size such
that $\mathrm{rank}(B_\Box) = d_{\mathrm{vir}}$, i.e., the virtual system
$G_{\mathrm{in} \Box}(\mathbf k)$ is fully mapped onto the physical space.
In this case we can use an SVD of $B_\Box = \mathbb{V}^\top \Sigma U$,
where $\mathbb{V}$ is an isometry, $\mathbb{V} \mathbb{V}^\top =
\mathbb{I}_{d_{\mathrm{vir}}}$, and $\Sigma$ is a diagonal strictly
positive matrix, to obtain from Eq.~\eqref{G_out} $\mathbb{V}
\left(G_{\mathrm{out}\Box}(\mathbf k) - A_\Box \right) \mathbb{V}^\top =
\Sigma U (D_\Box - G_{\mathrm{in} \Box}(\mathbf k))^{-1} U^\top \Sigma$,
which implies
\begin{align}
\det\left(\mathbb{V} \left(G_{\mathrm{out}\Box}(\mathbf k) - A_\Box \right) \mathbb{V}^\top\right)
= \frac{\det^2(\Sigma)}{\det(D_\Box - G_{\mathrm{in} \Box}(\mathbf
k))}\ .
\label{determinants}
\end{align}
Since all terms on the left-hand side are entries of CMs and thus bounded, it
follows that {$q_\Box(\mathbf k):= \det(D_\Box - G_{\mathrm{in}\Box}(\mathbf k))
\ge \delta>0$} [in particular, the parent Hamiltonian of the blocked GFPEPS
with {$\varepsilon(\mathbf k)= q_\Box(\mathbf k)$} in eq. \eqref{parent_Ham} is gapped and local].

It is now exactly this property which allows us to construct a gapped
interpolation from $G_{\mathrm{out}\Box} (\mathbf k)$ to the
topologically trivial state by adiabatically disentangling pairs of Majorana bonds (we can take $n_v$, $n_h$ to be even, since injectivity is stable under blocking) via
\begin{align}
\Gamma_{\mathrm{in}}^\varphi = \left(\begin{array}{cc}
\omega \sin\varphi & \mathbb{I}\cos\varphi\\
-\mathbb{I}\cos\varphi & \omega \sin\varphi
\end{array}\right)\ .
\end{align}
Here, $\Gamma_{\mathrm{in}}^\varphi$ is the CM of pairs of Majorana bonds
on horizontally or vertically adjacent sites, which for $\varphi = 0$
describes a maximally entangled state, corresponding to the initial
GFPEPS, while $\Gamma_\mathrm{in}^{\pi/2}$ corresponds to a product state
and thus, $G_{\mathrm{out}\Box}^{\pi/2}$ describes a topologically trivial
state~\cite{Note1}. Since from~\eqref{determinants}, $\det(D_\Box - G^\varphi_{\mathrm{in}\Box}(\mathbf k)) > 0$ for all
$\varphi \in [0, \tfrac{\pi}{2}]$, this interpolation corresponds
to a smooth gapped local Hamiltonian, showing that any injective GFPEPS is
in the trivial phase.

\begin{figure}[t]
\centering\includegraphics[width=\columnwidth]{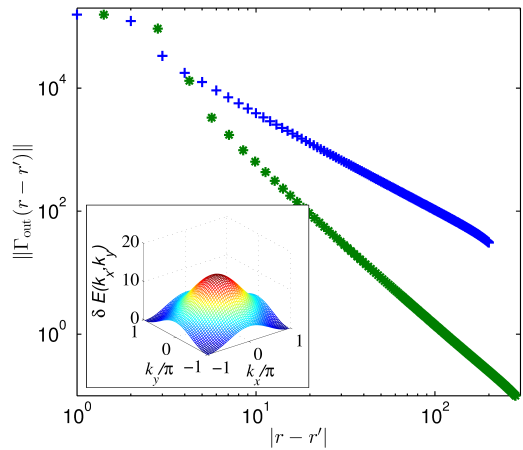}
\caption{Correlation functions as given by $\|  \Gamma_{\mathrm{out}}(\mathbf r - \mathbf r') \|_{\mathrm{tr}}$ for $\lambda = \frac{1}{\sqrt{2}}$ on a $500 \times 500$ lattice as a function of the distance $|\mathbf r - \mathbf r'|$ along the $x$ and $y$ directions (blue crosses, both lie on top of each other) and along $\hat x + \hat y$ (green stars). Insert:
Energy separation between the occupied and the unoccupied band as a function of $\mathbf{k}$.}
\label{analytical_gfpeps}
\end{figure}

This argument can be generalized to the case where
$\mathrm{rank}(B_\Box)<\mathrm{d_\mathrm{vir}}$ (i.e., the state is
non-injective), as long as $d_{\mathrm{vir}} \leq d_{\mathrm{ph}}$ and
{$q_\Box(\mb k) = \det(D_\Box - G_{\mathrm{in} \Box}(\mathbf k)) > 0$}.  In this case,
define $\Delta := \mathrm{min}_{\mathbf k} \det(D_\Box - G_{\mathrm{in} \Box
}(\mathbf k))$. It is always possible to rotate $M_\Box$ into ${M'_\Box}
= e^{- \epsilon Z} M_\Box e^{\epsilon Z}$ with some appropriate $Z = - Z^\top$ to
obtain $\mathrm{rank}(B_\Box) = d_{\mathrm{vir}}$, while keeping
{$\det(D'_\Box - G_{\mathrm{in} \Box}(\mathbf k)) > 0$} if $\epsilon$
is sufficiently small compared to $\Delta$. From there, it is again
possible to perform an adiabatic evolution to the trivial state as before.
If the initial GFPEPS was particle number conserving, this symmetry can be
kept along the path by using a particle-number conserving interpolation
$\Gamma_{\mathrm{in}}^\varphi$. Thus, our proof applies both to
topological insulators and topological superconductors.

\textit{Numerical results.---}%
We have performed numerical calculations on a $10 \times 10$ lattice for
the model
$H=\sum_{\mathbf{k}} (a_{\mathbf{k},\uparrow}^\dagger,
a_{\mathbf{k},\downarrow}^\dagger)
\,
(\mathbf{\bm\sigma}\cdot
\mathbf{d}(\mathbf{k}))
\,
(a_{\mathbf{k},\uparrow},
a_{\mathbf{k},\downarrow})^\top$,
with $\bm\sigma=(\sigma_1,\sigma_2,\sigma_3)$ the Pauli matrices, and
$\mathbf d (\mathbf k) = \left(\sin k_y,-\sin k_x,2 - \cos k_x - \cos k_y - e_S\right)$.
This model has Chern number $C = -1$ for $0 < e_S < 2$, $C = 1$ for $2 <
e_S < 4$ and $C = 0$ otherwise \cite{XLQi}.

First, we determined the minimal distance $\delta := \mathrm{max}_{\mathbf
k} \| G_{\mathrm{ex}}(\beta, \mathbf k) -
G_{\mathrm{GFPEPS}}(\mathbf k)\|_{\mathrm{tr}}$ between the
CM of $e^{-\beta H}/\tr(e^{-\beta H})$ (for $e_S=1$) at
$\beta = \infty$ and the one of the GFPEPS
$G_{\mathrm{GFPEPS}}(\mathbf{k})$ with a
given $\chi$. The results are shown in Fig.~\ref{numerics}a: We
find that the error, $\delta$, in the CM decreases exponentially
with the number of bond modes $\chi$. Since all physical quantities
depend solely on the CM, our results indicate that if
$\chi$ is increased, all relevant observables can be approximated by a
GFPEPS with exponentially decreasing error. Most importantly, the Hall
conductivity $- \frac{\sigma_{xy}}{2 \pi}$ reaches $C = -1$ with
exponentially decreasing difference, and the entropy of the
optimal GFPEPS approximation decreases exponentially with $\chi$.

\begin{figure}
\centering\includegraphics[width=\columnwidth]{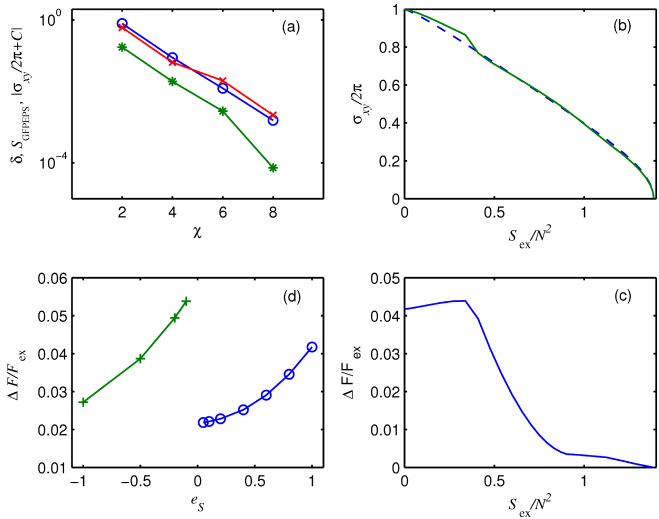}
\caption{(a) Error $\delta$ (see text) of the covariance matrix for $T = 0$, $e_S = 1$ (blue circles), entropy $S_{\mathrm{GFPEPS}}$ of the optimized GFPEPS (red crosses), difference between the Hall conductivity $-\frac{\sigma_{xy}}{2 \pi}$ and the Chern number $C = -1$ (green stars) for the optimized GFPEPS as a function of the number of Majorana modes $\chi$. 
(b) The blue dashed curve and solid green curve denote the Hall conductivity of the exact thermal state and
GFPEPS, respectively, as a function of the von-Neumann entropy of the exact state $S_{\mathrm{ex}}/N^2$. (c) Relative error of the free energy per site of the optimized
GFPEPS as a function of $S_{\mathrm{ex}}/N^2$. The entropies of the optimized GFPEPSs were roughly proportional to $S_{\mathrm{ex}}$. The Chern numbers of the Hamiltonians of which they are thermal states were always $C = -1$. (d) Relative error of the free energy per site of
the optimized GFPEPS as a function of the parameter $e_S$ of the exact
state at $T = 0$. The optimized GFPEPSs with Chern number $C = 0$ are displayed by green crosses and those with $C = -1$ by open blue circles.
} \label{numerics}
\end{figure}

We have also investigated the power of the GFPEPS to describe topological
insulators at finite temperature. This was done by minimizing the free energy
functional $F(\rho_{\mathrm{GFPEPS}}) = \tr(H \rho_{\mathrm{GFPEPS}}) - T
S(\rho_{\mathrm{GFPEPS}})$ ($S$: von Neumann entropy) of the above model at $e_S = 1$, with a GFPEPS
with $\chi=2$, i.e., one bond Majorana fermion per physical fermion, cf. Fig.~\ref{numerics}b,c:  For $T
\rightarrow 0$, the entropy of the GFPEPS does indeed converge to
zero, that is, it approaches a pure state; its analytical form is just the one
given in Eq.~\eqref{analytical_GFPEPS}, with $\lambda \approx 0.705$. This
shows that by minimizing the free energy as a function of $T$ one can
converge to pure states which are topological even for small $\chi$. Thus,
pure GFPEPS are well suited to describe topological insulators in numerical
simulations. We have substantiated this claim further by minimizing the free
energy $F(\rho_{\mathrm{GFPEPS}})$ for $\chi = 2$ as a function of $-1
\leq e_S \leq 1$ for $T = 0$, see Fig.~\ref{numerics}d. The entropies of the optimal GFPEPS were of the order of
$\leq 10^{-10}$ which is why their Chern numbers coincide with their Hall
conductivities, which jump from $0$ to $-1$ at $e_S \approx 0$. These results indicate that quantum phase transitions
can be detected by approximating the ground state energy with GFPEPS.

\emph{Conclusions.---}%
In this Letter, we have studied whether Projected Entangled Pair States
can be used to describe chiral topological states. We have answered this
question in the affirmative, by providing a class of Gaussian fermionic
PEPS (GFPEPS) describing systems with a non-zero Chern number; these
states can be ground states of either gapless strictly local Hamiltonians or gapped
Hamiltonians with algebraically decaying hoppings and/or pairings.  We have further
shown that the gaplessness of the strictly local parent Hamiltonian is a necessary
condition to have topological order.
Finally, we have numerically studied the ability of GFPEPS
to approximate chiral free fermion systems, and found that GFPEPS can
efficiently approximate both ground and thermal states of chiral
Hamiltonians with a small bond dimension, making them a well-suited tool
for the numerical study of chiral fermionic systems.

While we restricted our studies to Gaussian PEPS, it appears that one can
also describe interacting chiral systems with fermionic PEPS by twisting
the Gaussian PEPS projector $\mathcal E$ with a non-Gaussian map.  We
therefore believe that fermionic PEPS will also be suitable as a numerical
tool to study fractional quantum Hall systems.

After completion of this work, we learned that Dubail and Read had
independently obtained related results~\cite{Dub13}.

\textit{Acknowledgments.---}%
We thank the Benasque Center of Sciences, where part of this work has been
done, for their hospitality. T.\ B.\ W.\ acknowledges financial support by
the QCCC Elitenetzwerk Bayern.  N.\ S.\ acknowledges support by the
Alexander von Humboldt foundation. Part of the work has been supported by the EU Integrated Project SIQS.



\onecolumngrid
\appendix
\setcounter{equation}{0}
\newpage

\centerline{\textbf{Supplemental Material}}

\maketitle

\section{Free fermionic systems}

Throughout this paper we consider translationally invariant free fermion systems on an $N \times N$ square lattice. The free fermion Hamiltonians take the following general quadratic form:
\begin{align}
H = \sum_{\mb r, \mb r'} \sum_{j,m = 1}^f (T_{\mb r - \mb r'; j,m} a^{\dag}_{\mb r, j} a_{\mb r, m}
+ \Delta_{\mb r - \mb r'; j,m} a^{\dag}_{\mb r, j} a^{\dag}_{\mb r', m} + \bar{\Delta}_{\mb r - \mb r'; j,m} a_{\mb r', m} a_{\mb r, j}),
\label{free_ferm_ham}
\end{align}
where $\mb r$ and $\mb r'$ denote the site indices, $T_{\mb r - \mb r'}$ are Hermitian matrices, and $\bar{\Delta}_{\mb r - \mb r'; j,m}$ is the complex conjugate of $\Delta_{\mb r - \mb r'; j,m}$. Here $a_{\mb r, j}$ and $a_{\mb r,m}^\dg$ denote the fermionic annihilation and creation operators, respectively, whereas $j,m =1,\ \ldots, f$ is the index of orbitals.

In the presence of translational invariance, the Hamiltonian \eqref{free_ferm_ham} can be diagonalized by Fourier transforming to $\textbf{k}$-space.
If the pairing terms $\Delta_{\mb r - \mb r'}$ all vanish, i.e., the particle number is conserved, the Fourier transformed Hamiltonian can be written as
\begin{align}
H = \sum_{\mb k} \sum_{j,m = 1}^f a^\dg_{\mb k, j} h_{j,m}(\mb k) a_{\mb k, m}.
\label{free_ferm_ham_k}
\end{align}
Thus, the fermionic modes get completely decoupled by diagonalizing the $f
\times f$ matrix $h(\mb k)$. If the $\Delta_{\mb r - \mb r'}$ do not vanish,
the Hamiltonian contains pairing terms, which requires a Bogoliubov-de
Gennes transformation to diagonalize the Hamiltonian.

\section{Chern insulators}

The simplest chiral topological state appears in the integer quantum Hall effect, and has a gap in the bulk and gapless chiral edge states at the boundary. The lattice versions of that state are the so-called Chern insulators. The free fermion Hamiltonians of the Chern insulators are characterized by the Chern number, which is a topological invariant distinguishing different phases. Physically, the Chern number counts the number of chiral edge modes at the boundary and is related to the quantization of the Hall conductivity.

Let us suppose that the Hamiltonian \eqref{free_ferm_ham_k} has $n_{\mathrm{occ}}$ filled bands. Then, the Chern number is defined by \cite{TKNN-1982}
\begin{align}
C = \sum_{n = 1}^{n_{\mathrm{occ}}} \int_{\mathrm{BZ}} d^2 k \, F^{(n)}_{xy}(\mb k).
\label{ChernNumber}
\end{align}
Here $F^{(n)}_{xy}$ is the Berry curvature of the $n$th band, defined by $F^{(n)}_{xy} = \frac{\partial A^{(n)}_y}{\partial k_x} - \frac{\partial A^{(n)}_x}{\partial k_y}$, where the Berry connection $\mb A^{(n)}$ is given by $\mb A^{(n)} = - i \langle u^{(n)}(\mb k) | \bs \nabla | u^{(n)}(\mb k)\rangle$ and $| u^{(n)}(\mb k)\rangle$ are the Bloch functions for the $n$th band, i.e., the $n$th eigenvector of $h(\mb k)$ in Eq. \eqref{free_ferm_ham_k}.

\section{Fermionic Gaussian states}

Fermionic Gaussian states \cite{Bravyi-2005-smat}, i.e., ground and thermal states of quadratic Hamiltonians such as \eqref{free_ferm_ham}, are completely characterized by their two-point correlation functions due to Wick's theorem.
In order to understand the formalism, it is convenient to consider them in the basis of Majorana fermions defined by
\begin{align}
c_{2j - 1} = a_j^\dg + a_j, \ \ \ c_{2j} = -i(a_j^\dg - a_j),
\end{align}
($j =1, \ \dots, l$) which satisfy the Clifford algebra $\{ c_r, c_s\} = 2 \delta_{r,s} $. The two-point correlation function of a state $\rho$ is given by
\begin{align}
\Gamma_{r,s} = \frac{i}{2} \tr\left(\rho [c_r, c_s]\right)
\label{Cov_matrix}
\end{align}
where $\Gamma = - \Gamma^\top$.

An arbitrary operator can be expanded as
\begin{align}
X = \alpha \I + \sum_{p = 1}^{2 l} \sum_{1 \leq r_1 \leq ... \leq r_p \leq
2l} \alpha_{r_1, ..., r_p} c_{r_1} ... c_{r_p}\ .
\end{align}
Substituting $c_s$ by Grassmann variables $\theta_s$ satisfying
$\{ \theta_r, \theta_s\} = 0$, we obtain the Grassmann representation of $X$
\begin{align}
X = \alpha \I + \sum_{p = 1}^{2l} \sum_{1 \leq r_1 \leq ... \leq r_p \leq
2l} \alpha_{r_1, ..., r_p} \theta_{r_1} ... \theta_{r_p}\ .
\end{align}
In Grassmann representation, the density operator of any fermionic
Gaussian state takes the form \cite{Bravyi-2005-smat}
\begin{align}
\rho = \frac{1}{2^l} \exp\left(\frac{i}{2}\mb \theta^\top \Gamma \theta\right),
\end{align}
where the matrix \eqref{Cov_matrix} enters. For fermionic Gaussian states,
$\Gamma$ is called covariance (or correlation) matrix. In this case, we
have $\Gamma \Gamma^\top \leq \I$ with equality if and only if the state is
pure. For a pure Gaussian state $\rho$, a parent Hamiltonian $H$ which has
$\rho$ as its ground state can be constructed via
\begin{align} H = -i
\sum_{r,s=1}^{2 l} \Gamma_{r,s} c_r c_s.  \label{Parent_Ham} 
\end{align}

Gaussian linear maps map Gaussian states to Gaussian states. They are
defined via \cite{Bravyi-2005-smat}
\begin{align}
\label{eq:gaussianmap-integral}
X_{\mathrm{out}}(\mb \theta) = C \int \exp\left[S(\mb \theta, \mb \eta) + i \eta^\top \mb \mu\right] X_{\mathrm{in}}(\mu) D \eta D \mu,
\end{align}
where $C$ is a number and the action of the map is given by
\begin{align}
S(\mb \theta, \mb \eta) = \frac{i}{2} (\mb \theta^\top, \mb \eta^\top) \left(\begin{array}{cc}
A&B\\
-B^\top&D
\end{array}\right)
\left(\begin{array}{c}
\mb \theta\\
\mb \eta
\end{array}\right). \label{Gaussian_map}
\end{align}
Note that the Gaussian maps are also parametrized by a covariance matrix
\begin{align}
M = \left(\begin{array}{cc}
A&B\\
-B^\top&D
\end{array}\right) = - M^\top,
\end{align}
which is real and antisymmetric and fulfills $M M^\top \leq \I$. If $M M^\top = \I$, pure states are mapped to pure states.

Evaluating the Gaussian integrals in Eq. \eqref{eq:gaussianmap-integral} yields the covariance matrix of the output Gaussian state
as
\begin{align}
\Gamma_{\mathrm{out}} = B (D + \Gamma_{\mathrm{in}}^{-1}) B^\top + A \ .
\label{Gamma_out}
\end{align}

\section{Gaussian fermionic projected entangled-pair states}

Gaussian fermionic PEPS (GFPEPS) can be conveniently defined by using the above fermionic Gaussian state formalism.
In order to do so, we introduce $\chi $ maximally entangled virtual Majorana bonds between every two
neighboring sites of an $N \times N$ square lattice with periodic boundary conditions. They form the input Gaussian density operator, which has the following Grassmann representation:
\begin{align}
\rho_{\mathrm{in}}(\bs \mu) = \prod_{\mb r} \prod_{\alpha = 1}^\chi
\frac{1}{2^2} \exp\left[i (\mu^r_{\mb r, \alpha} \mu^l_{\mb r + \hat x,
\alpha} + \mu^u_{\mb r, \alpha} \mu^d_{\mb r + \hat y, \alpha})\right],
\label{rho_in}
\end{align}
where the products are taken over the sites located at $\mb r$ and the virtual indices $\alpha = 1, \ ..., \ \chi$ ($\chi$: number of bond Majorana modes).
The Majorana modes on each site $\mb r$ are mapped onto the space of physical fermions at $\mb r$ via
\begin{equation}
M = \left(\begin{array}{cc}
A&B\\
-B^\top&D
\end{array}\right)
\end{equation}
with $A \in \mathbb{R}^{2 f \times 2 f}$, $B \in \mathbb{R}^{2 f \times 4 \chi}$ and $D \in \mathbb{R}^{4 \chi \times 4 \chi}$ ($f$: number of physical orbitals per site).
Thus, the covariance matrix of the GfPEPS is
\begin{equation}
\Gamma_{\mathrm{out}} = B' (D' - \Gamma_{\mathrm{in}})^{-1} B'^\top + A',
\label{Gamma_gfpeps}
\end{equation}
where the primed matrices are defined via $M' = \bigoplus_{\mb r} M$ and we used $\Gamma_\mathrm{in}^2 = - \mathbb{I}$.

For the calculation of $\Gamma_{\mathrm{out}}$, we employ the formalism introduced in \cite{Kra09_appendix}, which we briefly recapitulate here:  $\Gamma_{\mathrm{in,out}}$ are circulant matrices and can thus be block-diagonalized by the unitary matrix representation $\mathcal{F}$ of the Fourier transform to reciprocal space. If we apply $\mathcal{F}$ from the left and $\mathcal{F}^\dg$ from the right on \eqref{Gamma_gfpeps}, we arrive at
\begin{equation}
G_{\mathrm{out}}(\mb k) = B (D - G_{\mathrm{in}}(\mb k))^{-1} B + A,
\end{equation}
where $G_{\mathrm{in,out}}(\mb k)$ are the blocks on the diagonals of $\mathcal{F} \Gamma_{\mathrm{in,out}} \mathcal{F}^\dg$, which are labeled by the reciprocal vector $\mb k$. They are the covariance matrices of the Fourier transformed physical Majorana modes $c_{\mb r,s}$ ($s = 1, \ldots, 2 f$)
\begin{equation}
d_{\mb k,s} = \frac{1}{N} \sum_{\mb r} e^{- i \mb r \cdot \mb k} c_{\mb r,s},
\end{equation}
and Fourier transformed virtual Majorana modes $\gamma_{\mb r, \alpha}^v$ ($v = l, r, u, d$ and $\alpha = 1, \ldots, \chi$)
\begin{equation}
\hat \gamma_{\mb k, \alpha}^v = \frac{1}{N} \sum_\mb{r} e^{-i \mb r \cdot \mb k} \gamma_{\mb r, \alpha}^v,
\end{equation}
respectively. Note that $d_{\mb k,s}$ are complex Majorana fermions lacking physical meaning. However, given  $G_{\mathrm{out}}(\mb k)$, the covariance matrix of physical Majorana modes in reciprocal space, $\Gamma_{\mathrm{out}}(\mb k)$ can be calculated via a unitary transformation as will be described below. Before, we would like to mention that
\begin{align}
\Gamma_{\mathrm{in}} &= \Gamma_h \oplus \Gamma_v,\\
\Gamma_{h/v} &= \bigoplus_{y/x=1}^N \left( \mathrm{Perm}(N,1,2,...,N-1) \otimes \left(\begin{array}{cc}
0&\I_{\chi}\\
0&0
\end{array}\right) - \mathrm{Perm}(2,3,...,N,1) \otimes \left(\begin{array}{cc}
0&0\\
\I_{\chi}&0
\end{array}\right) \right),
\end{align}
where $\mathrm{Perm}(i_1,i_2,...,i_N)$ is the permutation matrix with elements $\sum_{j = 1}^N \delta_{i_j,j}$ and $\I_n$ denotes the $n \times n$ identity matrix. Thus, we obtain
\begin{equation}
G_{\mathrm{in}} (\mb k) = \left(\begin{array}{cc}
0&\I_{\chi} e^{i k_x}\\
-\I_{\chi} e^{-i k_x}&0\\
\end{array}\right) 
\oplus
\left(\begin{array}{cc}
0&\I_{\chi} e^{i k_y}\\
-\I_{\chi} e^{-i k_y}&0\\
\end{array}
\right).
\end{equation}

If we write the inverse of $D - G_\mathrm{in}(\mb k)$ in terms of the adjugate matrix $\mathrm{Adj}(\cdot)$:
\begin{align}
[D - G_\mathrm{in}(\mb k)]^{-1} = \frac{\mathrm{Adj}(D - G_{\mathrm{in}}(\mb k))}{\det(D - G_{\mathrm{in}}(\mb k))},
\end{align}
it follows that the elements of $G_{\mathrm{out}}(\mb k)$ are fractions of finite-degree trigonometric polynomials,
\begin{equation}
[G_\mathrm{out}(\mb k)]_{s t} = \frac{p_{s t}(\mb k)}{q(\mb k)}.
\end{equation}
$p_{s t}(\mb k)$ and $q(\mb k) \equiv \det(D - G_\mathrm{in}(\mb k))$ are polynomials of $\sin k_{x}$, $\cos k_{x}$, $\sin k_{y}$, and $\cos k_{y}$ of degree $\leq 2 \chi$, bounded by the number of virtual Majorana bonds.
Note that $q(\mb k) \in \mathbb{R}$, since $D - G_{\mathrm{in}}(\mb k)$ is anti-Hermitian and has even dimensions, implying that its determinant is real. 

Now, the real, physical Majorana modes in $\mb k$-space can be obtained by taking the Fourier transform of the physical fermionic modes $a_{\mb r,j}$ in the relation
\begin{align}
c_{\mb r, 2 j -1} &= a_{\mb r,j}^\dg + a_{\mb r,j} \\
c_{\mb r, 2j} &= (-i)(a_{\mb r,j}^\dg - a_{\mb r,j}),
\end{align}
and employing the relations $a_{\mb k,j} = \frac{1}{2} ( c_{\mb k,2j-1} - i c_{\mb k,2j})$ and $a_{\mb k,j}^\dg = \frac{1}{2}(c_{\mb k,2j-1} + i c_{\mb k,2j})$. One obtains
\begin{equation}
\left(\begin{array}{c}
\hat d_{\mb k,2 j - 1}\\
\hat d_{\mb k,2j}\\
\hat d_{\mb k,2 j - 1}^\dg\\
\hat d_{\mb k,2j}^\dg
\end{array}\right)
= U 
\left(\begin{array}{c}
\hat c_{\mb k,2 j - 1}\\
\hat c_{\mb k,2j}\\
\hat c_{-\mb k,2 j - 1}\\
\hat c_{-\mb k,2j}
\end{array}\right),
\ \ 
U = U^\dg = \frac{1}{2}\left(\begin{array}{cccc}
\I_2 - i \omega&\I_2 + i \omega\\
\I_2 - i \omega& \I_2 + i \omega
\end{array}\right),
\end{equation}
where  $\omega = \left(\begin{array}{cc}
0&1\\
-1&0\end{array}\right)$.
The covariance matrix of the physical Majorana fermions in $\mb k$-space is given by
\begin{align}
\Gamma_{\mathrm{out}}(\mb k > 0) &= \left(\begin{array}{cc}
\Gamma_{++}(\mb k > 0)& \Gamma_{+-}(\mb k > 0)\\
\Gamma_{-+}(\mb k > 0)&\Gamma_{--}(\mb k > 0)
\end{array}\right) \\
&= \frac{1}{4}\left(\begin{array}{cc}
V&\overline V\\
\overline V&V
\end{array}\right)
\left(\begin{array}{cc}
G_{\mathrm{out}}(\mb k)&0\\
0&\overline G_{\mathrm{out}}(\mb k)
\end{array}\right)  \left(\begin{array}{cc}
V&\overline V\\
\overline V & V
\end{array}\right) \\
&= \frac{1}{2} \mathrm{Re} \left[ \left(\begin{array}{cc}
V \, G_{\mathrm{out}}(\mb k) \, V&V \, G_{\mathrm{out}}(\mb k) \, \overline V\\
\overline V \, G_{\mathrm{out}}(\mb k) \, V&\overline V \, G_{\mathrm{out}}(\mb k) \, \overline V
\end{array}\right) \right], \ \ V = \bigoplus_{j = 1}^{f} (\I_2 - i \omega). \label{Covariance_real}
\end{align}
Therefore, also 
\begin{align}
[\Gamma_{\mathrm{out}}]_{st}(\mb k) = \frac{\tilde p_{st}(\mb k)}{q(\mb k)}
\end{align}
is a fraction of trigonometric polynomials of degree $\leq 2 \chi$.

\section{Particle-number conserving Gaussian fPEPS}

Particle number conservation is equivalent to
\begin{align}
\frac{i}{2} \tr( \rho [a_{\mb r,j}^\dg, a_{\mb r',m}^\dg]) &= 0 \\
\frac{i}{2} \tr( \rho [a_{\mb r,j}, a_{\mb r',m}]) & = 0.
\end{align}
If we express this in terms of Majorana operators and perform a Fourier transformation to reciprocal space, we arrive at
\begin{equation}
[G_{\mathrm{out}}]_{2j-1,2m-1}(\mb k) \mp i \left([G_{\mathrm{out}}]_{2j,2m-1}(\mb k) + [G_{\mathrm{out}}]_{2j-1,2m}(\mb k)\right) - [G_{\mathrm{out}}]_{2j,2m}(\mb k) = 0.
\end{equation}
This is fulfilled, if and only if $G_{\mathrm{out}}(\mb k)$ can be written as 
$G_{\mathrm{out}}(\mb k) = \I_2 \otimes G_1(\mb k) + \omega \otimes G_2(\mb k)$, where $G_{1,2}(\mb k)$ are complex $f \times f$ matrices. Inserting this into \eqref{Covariance_real} yields
\begin{align}
\Gamma_{+-}(\mb k > 0) &= \Gamma_{-+}(\mb k > 0) = 0 \label{pm} \\
\Gamma_{++}(\mb k > 0) &= \Gamma_{--}(- \mb k) = \I_2 \otimes \mathrm{Re}(G_1(\mb k)) + \omega \otimes \mathrm{Re}(G_2(\mb k)) + \omega \otimes \mathrm{Im} (G_1(\mb k)) - \I_2 \otimes \mathrm{Im}(G_2(\mb k)). \label{pp}
\end{align}
Comparing this to
$G_1(\mb k) + i G_2(\mb k) = \mathrm{Re}(G_1(\mb k)) + i \mathrm{Re}(G_2(\mb k)) + i \mathrm{Im}(G_1(\mb k)) - \mathrm{Im}(G_2(\mb k))$, we note that the complex 
\begin{equation}
\Gamma_{\mathrm{out}}^{\mathbb{C}} (\mb k) := G_1(\mb k) + i G_2(\mb k)
\end{equation}
contains all the information about the correlation matrix in the case of particle number conservation. This property is due to the fact that $\I$, $\omega$ are the real representation of complex numbers.

In order ensure the GFPEPS has particle number conservation, it is best to impose the above symmetry already on the virtual level. This is done by grouping pairs of Majoranas together,
which in complex representation can be compactly written as 
\begin{equation}
\Gamma^{\mathbb{C}}_{\mathrm{in}}(\mb k) = \left(\begin{array}{cc}
0&\I_{\chi/2}e^{i k_x}\\
-\I_{\chi/2}e^{- i k_x}&0
\end{array}\right) \oplus 
\left(\begin{array}{cc}
0&\I_{\chi/2} e^{i k_y}\\
-\I_{\chi/2}e^{- i k_y}&0
\end{array}\right).
\end{equation}
The Gaussian map $M$ needs to preserve that symmetry, which is achieved by  $M = \I_2 \otimes M_1 + \omega \otimes M_2$, or in complex representation $M^{\mathbb{C}} = M_1 + i M_2$. Then
\begin{equation}
\Gamma^{\mathbb{C}}_{\mathrm{out}}(\mb k) = B^{\mathbb{C}} (D^{\mathbb{C}} - \Gamma_{\mathrm{in}}^{\mathbb{C}}(\mb k))^{-1} B^{\mathbb{C} \dg} + A^{\mathbb{C}}.
\end{equation}
Again $M^{\mathbb{C}} = -M^{\mathbb{C} \, \dg}$ and $M^{\mathbb{C}} M^{\mathbb{C} \, \dg} \leq \mathbb{I}_{f + 2 \chi}$ with equality for pure states. We note that particle number conservation of the output state can also be achieved by using normal virtual fermions for the bonds and applying a Gaussian map keeping the fermion number conservation symmetry.

\section{Gaussian fPEPS for Chern Insulators}

Let us consider a Chern insulator Hamiltonian with two orbitals on each site
\begin{align}
H = \sum_{\mb k} (a^\dg_{\mb k, \uparrow}, a^\dg_{\mb k,\downarrow}) h(\mb k)
\left(\begin{array}{c}
a_{\mb k, \uparrow}\\
a_{\mb k,\downarrow}
\end{array}\right)
\label{model_Ham}
\end{align}
with $h(\mb k) = \sum_{j = 1}^3 \sigma_j d_{j}(\mb k)$ leading to two
energy bands $\pm d(\mb k)$, where $d(\mathbf k) = |\mathbf d(\mathbf k)|$. We assume that it is gapped and
has one occupied band. In this two-orbital case, the Chern number in Eq.
\eqref{ChernNumber} can be simplified as \cite{XLQi-2006}
\begin{align}
C = \frac{1}{4 \pi} \int_{\mathrm{BZ}} d^2 k \, \hat {\mb d}(\mb k) \cdot \left(\frac{\partial \hat {\mb d}(\mb k)}{\partial k_x} \times \frac{\partial \hat {\mb d}(\mb k)}{\partial k_y}\right),
\label{Chern_from_d}
\end{align}
where $\hat {\mb d}(\mb k)$ is a three-component unit vector, $\hat {\mb
d}(\mb k) = \mb d (\mb k) / d(\mb k)$ and the integral runs over the first Brillouin zone (BZ).
For a given temperature $T=\frac{1}{\beta}$, the thermal density matrix for Eq. \eqref{model_Ham} is written as
\begin{align}
\rho(\beta) = \frac{e^{- \beta H}}{\tr(e^{-\beta H})} = \bigotimes_{\mb k} \frac{e^{- \beta H_{\mb k}}}{\tr(e^{-\beta H_{\mb k}})} := \bigotimes_{\mb k} \rho_{\mb k}(\beta) \label{rho_k},
\end{align}
The von Neumann entropy of this thermal state is given by
\begin{align}
S(\rho(\beta)) = \sum_{\mb k}S(\rho_{\mb k}(\beta)) = \sum_{\mb k} - \tr(\rho_{\mb k}(\beta) \ln \rho_{\mb k}(\beta)).
\end{align}
and the covariance matrix of $\rho(\beta)$ calculated via Eq. \eqref{Cov_matrix} is 
\begin{align}
\Gamma_{++}(\mb k) = \frac{\sinh(\beta d(\mb k))}{1 + \cosh(\beta d(\mb k))} \left(\begin{array}{cccc}
0&\hat d_3(\mb k)& \hat d_2(\mb k) & \hat d_1(\mb k)\\
-\hat d_3(\mb k)&0&-\hat d_1(\mb k)&\hat d_2(\mb k)\\
-\hat d_2(\mb k)& \hat d_1(\mb k)& 0 & -\hat d_3(\mb k)\\
-\hat d_1(\mb k)& -\hat d_2(\mb k)& \hat d_3(\mb k)&0
\end{array}\right)\ ,
\label{Gamma_thermal}
\end{align}
where due to particle number conservation the remaining parts of the covariance matrix are given by Eqs. \eqref{pm} and \eqref{pp}. $\Gamma_{++}(\mb k)$ could be written in complex representation; however it has one more symmetry, arising from the fact that 
the trace of the Hamiltonian is zero, $\tr(h(\mb k)) = 0$. Both symmetries can be incorporated simultaneously be representing $\Gamma_{++}(\mb k)$ by quaternions,
\begin{align}
\Gamma^{\mathbb{H}}(\mb k) = \frac{\sinh(\beta d(\mb k))}{1 + \cosh(\beta d(\mb k))} (i \hat d_3(\mb k) + j \hat d_2(\mb k) + k \hat d_1(\mb k)).
\end{align}
In order to see that, observe that the matrix representation of the quaternionic units is
\begin{equation}
\underline r = \left(\begin{array}{cccc}
1&0&0&0\\
0&1&0&0\\
0&0&1&0\\
0&0&0&1
\end{array}\right), \
\underline i = \left(\begin{array}{cccc}
0&1&0&0\\
-1&0&0&0\\
0&0&0&-1\\
0&0&1&0
\end{array}\right), \
\underline j = \left(\begin{array}{cccc}
0&0&1&0\\
0&0&0&1\\
-1&0&0&0\\
0&-1&0&0
\end{array}\right), \
\underline k = \left(\begin{array}{cccc}
0&0&0&1\\
0&0&-1&0\\
0&1&0&0\\
-1&0&0&0
\end{array}\right),
\end{equation}
where $\underline r$ denotes the real unit. 

Again, the desired symmetry of the output state is obtained by imposing it on the virtual level and ensuring that the Gaussian map keeps it. We realize that the virtual modes already have the desired symmetry, since their covariance matrix $\Gamma_{\mathrm{in}}^{\mathbb{C}}$ after regrouping the virtual indices can be written directly in quaternionic representation as
\begin{align}
\Gamma^{\mathbb{H}}_{\mathrm{in}}(\mb k) = \left(\begin{array}{cc}
\I_{\chi/2} (j \cos(k_x) + k \sin(k_x))&0 \\
0& \I_{\chi/2} (j \cos(k_y) + k \sin(k_y))
\end{array}\right),
\end{align}
because the representation of a quaternion $q = a + i b + j c + k d$ by complex matrices is
$\underline q =  \left(\begin{array}{cc}
a + i b& c + i d\\
-c + i d& a - i b
\end{array}\right)$. The Gaussian map keeps the symmetry by requiring that $M$ can also be represented by quaternions, that is
\begin{equation}
M^{\mathbb{H}} = M_r + i M_i + j M_j + k M_k \in \mathbb{H}^{(1 + \chi) \times (1 + \chi)}
\end{equation}
(note that $\chi$ has to be even). Then, the output covariance matrix is simply
\begin{align}
\Gamma_{\mathrm{out}}^\mathbb{H}(\mb k) = B^\mathbb{H} (D^\mathbb{H} - \Gamma_{\mathrm{in}}^\mathbb{H}(\mb k))^{-1} B^{\mathbb{H}\, \dg} + A^\mathbb{H}\ ,
\end{align}
where $M^\mathbb{H} = -M^{\mathbb{H}\,\dg}$ and $M^\mathbb{H} M^{\mathbb{H}\,\dg} \leq \mathbb{I}_{1 + \chi}$ with equality for pure states. The quaternionic representation has been used in the numerical calculations of this work in order to exploit the symmetries of the Hamiltonian\cite{quat_M}.

Given a mixed or pure GFPEPS (e.g., obtained after performing a minimization of the free energy), its Hall conductivity can be calculated by setting
\begin{align}
\Gamma_{\mathrm{out}}^{\mathbb{H}}(\mb k) = \frac{\sinh(\beta d(\mb k))}{1 + \cosh(\beta d(\mb k))} (i \hat d_3(\mb k) + j \hat d_2(\mb k) + k \hat d_1(\mb k)).
\end{align}
to extract $\hat {\mb d}(\mb k)$ and $\beta d(\mb k)$. The $\mb d$-vector from the GFPEPS gives rise to the Hall conductivity \cite{XLQi-2006}
\begin{align}
- \frac{\sigma_{xy}}{2 \pi} = \frac{1}{4 \pi} \int_{\mathrm{BZ}} d^2 k \, f(\beta, \mb k) \, \hat {\mb d}(\mb k) \cdot \left(\frac{\partial \hat {\mb d}(\mb k)}{\partial k_x} \times \frac{\partial \hat {\mb d}(\mb k)}{\partial k_y}\right) \, ,
\end{align}
where $f(\beta, \mb k) = \frac{1}{e^{-\beta d(\mb k)}+1} - \frac{1}{e^{\beta d(\mb k)} + 1}$.


\begin{thebibliography}{99}


\bibitem{Ver04}
F. Verstraete and J. I. Cirac, cond-mat/0407066; F. Verstraete and J. I. Cirac, Phys. Rev. A {\bf 70}, 060302 (2004).

\bibitem{Has05}
M. B. Hastings, Phys. Rev. B {\bf 73}, 085115 (2006).

\bibitem{Wol08}
M. M. Wolf, F. Verstraete, M. B. Hastings, and J. I. Cirac,
Phys. Rev. Lett. {\bf 100}, 070502 (2008).

\bibitem{Pol09}
F. Pollmann, A. M. Turner, E. Berg, and M. Oshikawa,
Phys. Rev. B {\bf 81}, 064439 (2010).

\bibitem{Che11}
X. Chen, Z.-C. Gu, and X.-G. Wen,
Phys. Rev. B {\bf 83}, 035107 (2011).

\bibitem{Sch11}
N. Schuch, D. Perez-Garcia, and I. Cirac,
Phys. Rev. B {\bf 84}, 165139 (2011).

\bibitem{Kit03}
A. Kitaev, Ann. Phys. {\bf 303}, 2 (2003).

\bibitem{And73}
P. W. Anderson, Mater. Res. Bull. {\bf 8}, 153 (1973).

\bibitem{Lev05}
M. A. Levin and X.-G. Wen,
Phys. Rev. B {\bf 71}, 045110 (2005).

\bibitem{Ver06}
F. Verstraete, M. M. Wolf, D. Perez-Garcia, and J. I. Cirac,
Phys. Rev. Lett. {\bf 96}, 220601 (2006).

\bibitem{Sch10}
N. Schuch, J. I. Cirac, and D. Perez-Garcia, Ann. Phys. {\bf 325}, 2153 (2010).

\bibitem{Bue08}
O. Buerschaper, M. Aguado, and G. Vidal, Phys. Rev. B {\bf 79}, 085119 (2009).

\bibitem{Gu08}
Z.-C. Gu, M.~Levin, B. Swingle, and X.-G. Wen, Phys. Rev. B {\bf 79}, 085118 (2009).

\bibitem{Note_Ber11}
Note however that in \cite{Ber11} it has been shown
that expectation values of observables in chiral models can be expressed as a
contraction of tensor networks, although the question whether states can be
expressed as tensor networks was left open.

\bibitem{Ber11}
B. B\'{e}ri and N. R. Cooper, Phys. Rev. Lett. {\bf 106}, 156401 (2011).

\bibitem{TopIns}
F. D. M. Haldane, Phys. Rev. Lett. {\bf 61}, 2015 (1988); M. Z. Hasan and C. L. Kane, Rev. Mod. Phys.
{\bf 82}, 3045 (2010); X.-L. Qi and S.-C. Zhang, Rev. Mod. Phys. {\bf 83}, 1057 (2011).

\bibitem{Per10}
D. Perez-Garcia, M. Sanz, C. E. Gonzalez-Guillen, M. M. Wolf, and J. I. Cirac,
New J. Phys. {\bf 12} 025010 (2010).

\bibitem{Cir09}
J. I. Cirac and F. Verstraete, J. Phys. A: Math. Theor. {\bf 42}, 504004 (2009).

\bibitem{Bro07}
C. Brouder, G. Panati, M. Calandra, C. Mourougane, and N. Marzari,
Phys. Rev. Lett. {\bf 98}, 046402 (2007).

\bibitem{Mar12}
N. Marzari, A. A. Mostofi, J. R. Yates, I. Souza, and D. Vanderbilt, Rev. Mod. Phys. {\bf 84}, 1419 (2012).

\bibitem{Kra09}
C. V. Kraus, N. Schuch, F. Verstraete, and J. I. Cirac,
Phys. Rev. A {\bf 81}, 052338 (2010).

\bibitem{Per08}
D.~Perez-Garcia, F.~Verstraete, J.~I. Cirac, and M.~M. Wolf,
\newblock Quantum Inf. Comput. {\bf 8}, 0650 (2008), arXiv:0707.2260.

\bibitem{Bravyi-2005} S. Bravyi, Quantum Inf. Comput. \textbf{5}, 216 (2005).

\bibitem{Supp}
See Supplemental Material for an introduction on free fermionic systems and Chern insulators, a detailed explanation of the Gaussian formalism and GFPEPS, GFPEPS with particle number conservation, and how the Chern number can be extracted from such a GFPEPS.

\bibitem{Note1}
Note that this construction does not work in 1D if the
number of Majorana bonds is odd, such as for  the Kitaev
chain~\cite{Kit00}.

\bibitem{Kit00}
A. Kitaev, Phys. Usp. \textbf{44}, 131 (2001).

\bibitem{XLQi} X.-L. Qi, Y.-S. Wu, and S.-C. Zhang, Phys. Rev. B \textbf{74}, 085308 (2006).

\bibitem{Dub13}
J.~Dubail and N. Read, arXiv:1307.7726.



\end{thebibliography}

\begin{thebibliography}{99}

\setcounter{NAT@ctr}{28}

\bibitem{TKNN-1982} D. J. Thouless, M. Kohmoto, M. P. Nightingale, and M. den Nijs, Phys. Rev. Lett. \textbf{49}, 405 (1982).

\bibitem{Bravyi-2005-smat} S. Bravyi, Quantum Inf. Comput. \textbf{5}, 216 (2005).

\bibitem{Kra09_appendix}
C. V. Kraus, N. Schuch, F. Verstraete, and J. I. Cirac,
Phys. Rev. A {\bf 81}, 052338 (2010).

\bibitem{XLQi-2006} X.-L. Qi, Y.-S. Wu, and S.-C. Zhang, Phys. Rev. B \textbf{74}, 085308 (2006).

\bibitem{quat_M}
Note that the topological analytical GFPEPS defined in the main body of the paper is not given in the standard real representation for quaternions, but related by a basis change to the one
for 
\begin{align*}
\Gamma_{\mathrm{in}}^{\mathbb{H}}(\mb k) = \left(\begin{array}{cc}
i \cos(k_x) + k \sin(k_x)& 0\\
0& i \cos(k_y) + k \sin(k_y)
\end{array}\right)
\end{align*}
and
\begin{align*}
M^\mathbb{H} = \left(\begin{array}{ccc}
i \, a_1&j \, a_2 - k \,a_2&-k \, a_3\\
j \, a_2 - k \, a_2&i \, a_4& -a_5 + i \,a_5\\
-k \, a_3&a_5+i \, a_5&i \,a_4\\
\end{array}\right)
\end{align*}
with $a_1 = -1 + 2 \lambda$, $a_2 = \sqrt{\lambda - \lambda^2}$, $a_3 = \sqrt{2(\lambda - \lambda^2)}$, $a_4 = -1 + \lambda$, $a_5 =\frac{1}{\sqrt 2} \lambda$.



\end{thebibliography}
\end{document}